\documentclass[12pt]{article}
\input epsf

\makeatletter

\@addtoreset{equation}{section}
\makeatother

\def\balpha{{\mbox{\boldmath $\alpha$}}}
\def\bbeta{{\mbox{\boldmath $\beta$}}}              
\def\bgamma{{\mbox{\boldmath $\gamma$}}}
\def\la{\mathrel{\mathpalette\fun <}}
\def\ga{\mathrel{\mathpalette\fun >}}
\def\fun#1#2{\lower3.6pt\vbox{\baselineskip0pt\lineskip.9pt
  \ialign{$\mathsurround=0pt#1\hfil##\hfil$\crcr#2\crcr\sim\crcr}}}

\begin{document}

\title{{\vskip -80pt
  \begin{flushright}
   {\normalsize{CU-TP-982\\ \vskip-10pt
    hep-th/0007038}}
  \end{flushright}
  \vskip 15pt}
Duality and Massless Monopoles\footnote{Talk delivered at Quarks
 2000, Pushkin 
Russia, May 2000}}
\author{Erick J. Weinberg \\
   Department of Physics, Columbia University \\
   New York, NY 10027, USA}
\date{}
\maketitle
\begin{abstract}
Duality arguments suggest the existence of massless magnetic
mono\-poles in gauge theories with the symmetry broken to a
non-Abelian subgroup.  I discuss how these arise and show how they are
manifested as clouds of massless fields surrounding massive monopoles.
The dynamics of these clouds is discussed, and the scattering of
massless monopole clouds and massive monopoles is described.
\end{abstract}

\section{Introduction}

A quarter century ago 't Hooft and Polyakov showed \cite{thooft} that
magnetic mono\-poles could arise as topologically stable classical
solutions in certain spontaneously broken gauge theories.  The fact that
these included all grand unified theories led to a considerable
theoretical effort to study the detailed properties and the
astrophysical and cosmological implications of these GUT
monopoles, as well as to a number of experimental searches.  However, the
failure of these searches to detect any monopoles, together with
astrophysical arguments that place stringent bounds on the monopole flux
at the Earth's surface, has quite understandably led to a decreased
interest from the phenomenological point of view.

Nevertheless, these monopole solutions remain valuable as tools
for probing the properties of quantum field theory.  In particular,
the properties of monopoles in the Bogomol'nyi-Prasad-Sommerfield (BPS)
limit \cite{bps} has led to the conjecture \cite{montonen} that the
electric-magnetic duality of Maxwell's equations might find a quantum
field theoretic generalization that exchanges magnetically charged
solitons with electrically charged elementary particles.  The prime
candidate \cite{osborn}  for this is $N=4$ supersymmetric Yang-Mills
theory.  In the case of SU(2) broken to U(1), the spectrum of massive
particles is invariant under the simultaneous interchange of weak and
strong coupling  and of electric and magnetic charges.  This result
generalizes fairly easily to larger gauge groups, as long as the
subgroup that remains unbroken is Abelian.  However, if the unbroken
subgroup is Abelian, the elementary particle sector contains massless
particles with electric-type charge (the ``gluons'' and their
superpartners).  Duality would then require the existence of massless
magnetically charged particles.  However, one can easily show that
these theories have no massless classical soliton solutions.  In this
talk, I will argue that, nevertheless, one can find evidence for the
required massless monopoles by studying the dynamics of the massive
monopoles.

\section{BPS monopoles in SU(2) and larger groups}

Throughout this talk I will work in the BPS limit, with monopoles
obeying 
\begin{equation}
    B_i = D_i\Phi  \, .
\label{bpseq}
\end{equation}
While this limit can be obtained by taking a limit of coupling
constants, it is most naturally understood in the context of
Yang-Mills theory with extended supersymmetry, where Eq.~(\ref{bpseq})
is equivalent at the classical level to requiring that the soliton
preserve half of the supersymmetry.\footnote{Although I have $N=4$
supersymmetric Yang-Mills in mind, the only feature of this theory
that I will use explicitly is the fact that the Higgs field is in the
adjoint representation.  Note also that there is no confinement in
this theory, even if there is an unbroken non-Abelian subgroup.}  This
ensures that the BPS mass formula
\begin{equation}
    M = Q_M v =  \int dS_i {\rm Tr}\, B_i\Phi 
\label{bpsmass}
\end{equation}
is preserved by quantum corrections. 

One important feature of the BPS limit is that the Higgs scalar
becomes massless and so can mediate a long-range attractive force
between two monopoles.  It turns out that this can exactly balance the
magnetic repulsion between static monopoles, thus allowing static
multimonopole solutions to exist.  In fact, for the case of SU(2)
broken to U(1),  all higher charged BPS solutions can
be understood as multimonopole solutions of this sort.  Not only does
a solution with $n$ units of magnetic charge have $n$ times the mass
of the unit monopole, but index theory methods show \cite{su2index}
that the number collective coordinates needed to quantize the solution
is precisely $4n$, corresponding to a three-dimensional position and a
U(1) phase for each of the $n$ constituent monopoles; excitation of these
phase coordinates gives independent dyonic electric charges to the
individual monopoles.

In order to have the possibility of symmetry breaking to a non-Abelian
subgroup, we must start with a gauge group of rank $r \ge 2$.  The
generators of this gauge group can be chosen to be a set of $r$
commuting $H_i$ that form a basis for the Cartan subalgebra, together
with a set of raising and lowering operators associated with the roots
$\balpha$.  The asymptotic adjoint Higgs field in some fixed direction
can always be brought into the form
\begin{equation}
      \Phi_0 = {\bf h} \cdot {\bf H}  \, .
\end{equation}
If the $r$-component vector $\bf h$ has nonzero inner products with
all of the roots, then the gauge symmetry is broken maximally, to
U(1)$^r$; if not, then the roots orthogonal to $\bf h$ yield the root
diagram of an unbroken non-Abelian group $K$.

The long range part of the magnetic field must lie in the unbroken
part of the gauge group.  Hence, in the direction used to define
$\Phi_0$, the leading part of asymptotic magnetic field can also be
brought into the Cartan subalgebra and written in the form
\begin{equation}
      B_i = {\bf g} \cdot {\bf H} {{\hat r}_i\over r^2}  \, .
\end{equation}

In the case of maximal symmetry breaking there are $r$ topological
charges, one for each of the unbroken U(1) factors.  The connection
between these and the magnetic charge is most easily seen by choosing
a set of $r$ simple roots $\bbeta_a$. These form a basis for the root
lattice of the Lie group with the property that any root can be written as
a linear combination of simple roots with coefficients all of the same
sign.  There are many possible choices for the simple roots, but the
vector $\bf h$ associated with the Higgs field can be used to pick out
a unique set satisfying ${\bf h} \cdot {\bbeta_a} > 0$.  The
topological quantization condition \cite{quantization} then takes the
form
\begin{equation}
      {\bf g} = {4\pi\over e} \sum_a n_a {\bbeta_a \over \bbeta_a^2} 
\label{gdef}
\end{equation}
where the integers $n_a$ are the topological charges.  The BPS mass
formula, Eq.~(\ref{bpsmass}), can be written as 
\begin{equation}
      M = {\bf g}\cdot {\bf h} 
     = \sum_a n_a \left( {4\pi\over e} {\bf h} \cdot \bbeta_a  \right)
     \equiv \sum_a n_a m_a  \, ,
\label{msbmass}
\end{equation} 
while the number of collective coordinates is \cite{Gindex}
\begin{equation}
   {\cal N}= 4\sum_a n_a  \, .
\label{msbindex}
\end{equation}

These results suggest that, just as in the SU(2) case, all higher
charged solutions should be viewed as multimonopole configurations.
Now, however, there are $r$ species of fundamental monopoles, one for
each U(1) factor, with the $a$th species having mass $m_a$,
topological charges $n_b=\delta_{ab}$, and four degrees of freedom
(three for center-of-mass motion and one U(1) phase).  These
fundamental monopoles can be explicitly obtained by embedding the
SU(2) unit monopole in the SU(2) subgroups defined by the various
$\bbeta_a$.

When the symmetry breaking is nonmaximal, to $K \times {\rm
U(1)}^{r-k}$, some of the simple roots, which I denote by $\bgamma_i$,
are orthogonal to $\bf h$ and form a complete set of simple roots for
$K$.  The remainder, denoted $\tilde \bbeta_a$, can required, as
before, to obey ${\bf h} \cdot {\tilde\bbeta_a} > 0$.
Equation~(\ref{gdef}) is replaced by
\begin{equation}
      {\bf g} = {4\pi\over e} \sum_a \tilde n_a {\tilde \bbeta_a 
              \over \tilde \bbeta_a^2} 
          + \sum_i q_i{\bgamma_j \over \bgamma_j^2}  \, .
\label{NUSgdef}
\end{equation}
The $\tilde n_a$ and $q_i$ are integers, with the former being the
conserved topological charges.\footnote{In contrast with the maximally
broken case, the simple roots are not uniquely determined by the
requirement that their inner product with $\bf h$ be positive.  The
various allowed sets are related by gauge transformations of the
unbroken group.  The $\tilde n_a$ are invariant under these
transformations, but the $q_i$ are not.}  

In general, the corresponding magnetic field has
both Abelian and non-Abelian components.  In order to avoid certain
pathologies \cite{pathologies} associated with non-Abelian magnetic
charges, I will assume for the remainder of this talk that ${\bf g}
\cdot \bgamma_i =$ for all $i$, so that the long-range magnetic field
is purely Abelian.  (There is little loss of generality in this
assumption since, given a configuration with nonzero ${\bf g} \cdot
\bgamma_i$, the additional monopoles needed to cancel the non-Abelian
part of the total magnetic charge can be placed at an arbitrarily
large distance.)  With this assumption,
Eqs.~(\ref{msbmass}) and (\ref{msbindex}) are replaced by
\begin{equation}
      M =  \sum_a \tilde n_a m_a 
\end{equation} 
and \cite{nonabelindex}
\begin{equation}
   {\cal N}=  4\sum_a\tilde n_a + 4 \sum_i q_i \, .
\end{equation} 

Analogy with the maximally broken case would then suggest that there
is one fundamental monopole, with mass $m_a$ and four degrees of
freedom, associated each of the $\tilde\bbeta_a$, and one massless
monopole, also with four degrees of freedom, associated with each of
the $\bgamma_i$.  Indeed, the embedding construction for the massive
fundamental monopoles goes through pretty much as before.  The massless
monopoles, on the other hand, cannot be constructed in this manner:
applying the embedding construction to the SU(2) subgroup
corresponding to one of the $\bgamma_i$ simply yields a pure vacuum
solution.  In fact, it is easy to show that there are no
localized classical solutions with zero energy.

\section{Low energy monopole dynamics}

Although we cannot obtain classical solutions corresponding to
isolated massless monopoles, one can find evidence of these monopoles in the
dynamics of multimonopole systems.  The moduli space approximation
\cite{msa} is a convenient tool for studying such systems at low
energy.  The essential idea is to approximate solutions with slowly
moving monopoles\footnote{These includes dyons with small electric
charges, since these correspond to slowly varying U(1) phases.} as
being motion on the moduli space of static BPS multimonopole
solutions.  More precisely, let \{$A_i^{\rm BPS}({\bf r}, z)$,
$\Phi^{\rm BPS}({\bf r}, z)$\} be a family of gauge-inequivalent BPS
solutions parameterized by a set of collective coordinates $z_a$.  In
the moduli space approximation one adopts the Ansatz
\begin{eqnarray}
     A_0({\bf r},t) &=& 0 \cr 
     A_i({\bf r},t) &=& U^{-1}({\bf r},t)  A_i^{\rm BPS}({\bf r},
     z(t)) U({\bf 
       r},t) -{i} U^{-1}({\bf r},t) \partial_i U({\bf r},t) \cr
       \Phi({\bf r},t) &=& U^{-1}({\bf r},t) \Phi^{\rm BPS}({\bf r}, z(t))
      U({\bf r},t) \, .
\label{modspaceansatz}
\end{eqnarray}
With this Ansatz, the time derivatives of the fields are of the form
\begin{eqnarray}
    \dot A_i &=& \dot z_j \left[{\partial A_i \over \partial z_j} + D_i
           \epsilon_j \right]  \equiv \dot z_j  \delta_j A_i  \cr
    \dot \Phi &=& \dot z_j \left[{\partial \Phi \over \partial z_j} + [\Phi,
           \epsilon_j] \right]  \equiv \dot z_j  \delta_j \Phi   \, .
\end{eqnarray}
The terms involving $\epsilon_j$ arise from differentiating the
gauge function $U({\bf r},t)$; they are fixed uniquely by Gauss's
law, which turns out to be equivalent to imposing a background gauge
condition on $\dot A_i$ and $\dot \Phi$.  

Substituting this Ansatz into the Yang-Mills Lagrangian gives
\begin{equation}
    L_{MS} = {1\over 2} \int d^3r {\rm Tr}\left[ {\dot A_i}^2 + {\dot
          \Phi}^2 + B_i^2 + 
          D_i\Phi^2 \right] 
   = {1\over 2} g_{ij}(z) \dot z_i \dot z_j  + M
\label{LMS}
\end{equation}
where the static energy $M$ is constant on the moduli space and
\begin{equation}
    g_{ij}(z) = \int d^3r \left[ \delta_i A_k \delta_j A_k + 
        \delta_i \Phi  \delta_j \Phi  \right]  \, .
\label{metricdef}
\end{equation}
If we interpret $g_{ij}$ as a metric on the moduli space, then the
solutions to $ L_{MS}$ are simply geodesic motion on the moduli space.  In most
cases, it turns out not to be practicable to use Eq.~(\ref{metricdef})
directly to determine $g_{ij}(z)$.  However, by using indirect methods
the moduli space metrics for a number of interesting cases have been
by more indirect methods. I will make use of some of these results in
the next two sections.

\section{An SO(5) example}

The simplest example where one finds evidence of massless monopoles
arises with the gauge group SO(5), whose root diagram is shown in
Fig.~1.  With the Higgs field vector $\bf h$ is in the direction shown
on the left, the symmetry is broken to U(1)$\times$U(1).  There are
two species of massive fundamental monopoles, corresponding to the
simple roots $\bbeta$ and $\bgamma$.  If instead the Higgs vector is
vertical, as shown on the right, the unbroken gauge group is
SU(2)$\times$U(1), with the SU(2) having roots $\pm \bgamma$.  The
fundamental $\bbeta$-monopole remains massive, but now $\bgamma$
corresponds to an elusive massless monopole.

I will focus on solutions with
\begin{equation}
   {\bf g} = {4\pi\over e} \left( {\bbeta\over \bbeta^2} 
        + {\bgamma\over \bgamma^2} \right) \, .
\label{so5g}
\end{equation}
In the maximally broken case, these are composed of two distinct
massive fundamental monopoles, and have an eight-dimensional moduli
space whose metric is \cite{kwy} \pagebreak

\vskip 5mm
\begin{center}
\leavevmode
\epsfysize =2in\epsfbox{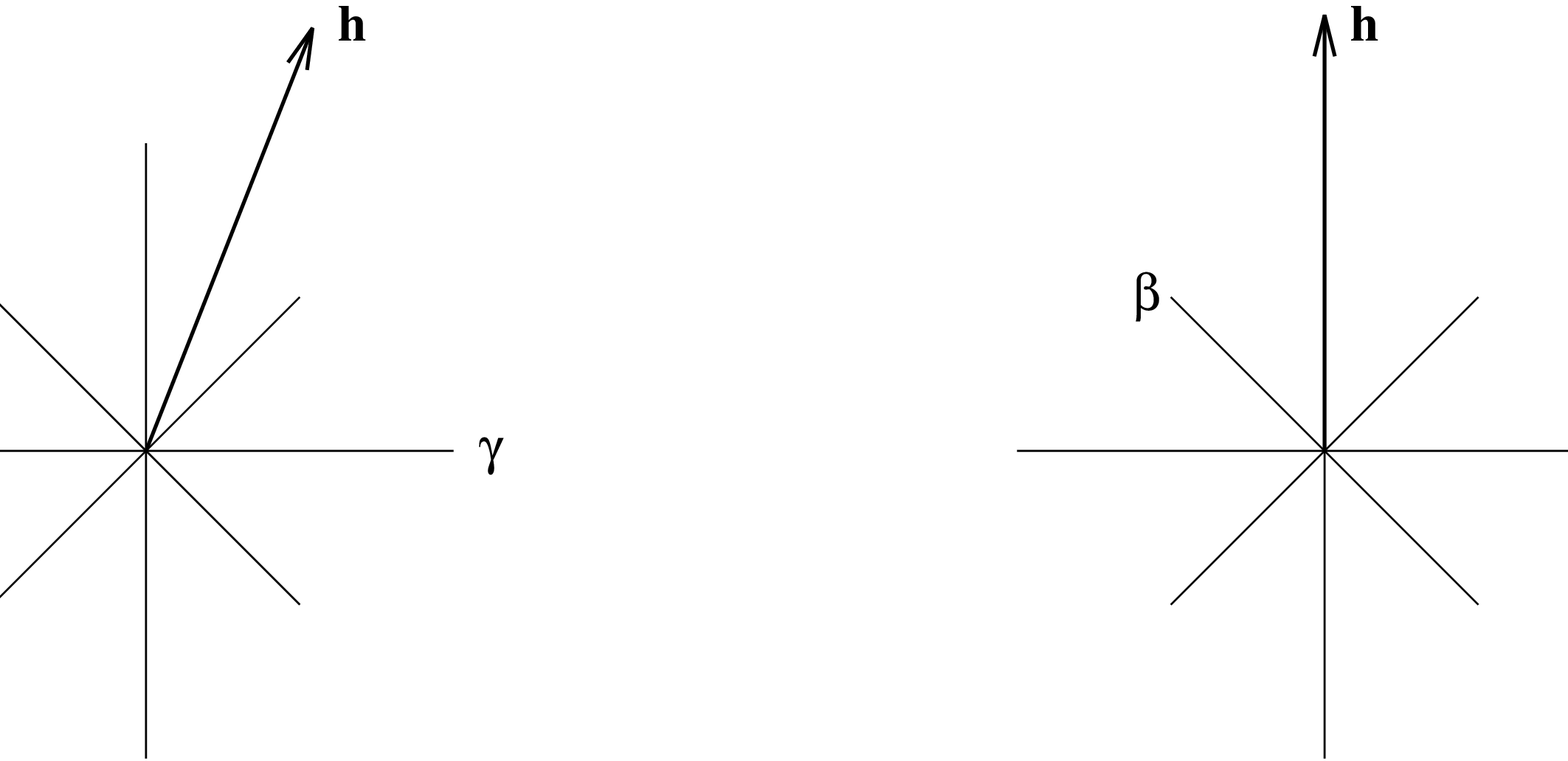}
\end{center}
\vskip 8mm
\begin{quote}
{\bf Figure 1:} {\small
The root diagram of $SO(5)$. With the Higgs vector $\bf h$ oriented as
in (a) the gauge symmetry is broken to U(1)$\times$U(1), while with
the orientation in (b) the breaking is to SU(2)$\times$U(1). }
\end{quote}
\vskip 0.40cm

\begin{eqnarray}
    ds^2 &=& M d{\bf X}_{\rm cm}^2 +  {16\pi^2 \over M
    }d\chi_{\rm 
    tot}^2 +  \left(\mu +{k\over r}\right) \left[ dr^2 +r^2(d\theta^2
    + \sin^2\theta d\phi^2) \right]  \cr & & \qquad 
  + k^2 \left(\mu +{k\over r}\right)^{-1} (d\psi +d\cos\theta d\phi)^2  \, .
\label{MSBmetric}
\end{eqnarray}
Here ${\bf X}_{\rm cm}$ is the center-of-mass position of the two
monopoles; $r$, $\theta$, 
and $\phi$ are their relative coordinates; and $\chi_{\rm  tot}$ and
$\psi$ are overall and relative U(1) phases.  $M$ and $\mu$ are the
total and reduced masses of the monopole pair, and $k$ is a numerical
constant related to the normalization of roots.

With SU(2)$\times$U(1) breaking, Eq.~(\ref{so5g}) corresponds to a
combination of one massive and one massless monopole.  It turns out to
be relatively straightforward to solve the field equations explicitly
\cite{so5}.  The solutions are spherically symmetric and depend on eight
parameters: three position coordinates {\bf X}, four SU(2)$\times$U(1) phase
angles $\alpha$, $\beta$, $\gamma$, and $\chi$. and one last
parameter, $a$, that can take on any real positive value and whose
interpretation will become clear shortly.  The fields can be
decomposed according their transformations under the unbroken SU(2).
The singlet terms are just what would be obtained by embedding the
unit SU(2) monopole using the composite root $2\bbeta+\bgamma$; they
have no dependence on $a$.  By themselves, they would describe an
object with a massive core surrounded by a simple Coulomb magnetic
field.  The doublet terms fall exponentially fast outside the monopole
core and are relatively uninteresting.  Finally, the triplet parts of
the gauge and Higgs fields are given by invariant tensors multiplied
by the function
\begin{equation}
   G(r, a) = \left[{v \over \sinh (evr)} -{1 \over er} \right]
          \left[ 1 +  {r\over a} \coth(vr/2) \right]^{-1}  \, .
\end{equation}
For $r \la a$, the second factor on the right hand side is
approximately unity, so $G$, and hence the corresponding $A_i$, falls
as $1/r$.  This produces a Coulomb magnetic field in the unbroken SU(2)
subgroup.  However, for $r \ga a$, the last factor gives an additional
$1/r$, so the triplet part of $A_i$ falls as $1/r^2$, and the
non-Abelian component of the magnetic field fall more rapidly.

Hence, we may view the solution as being composed of a massive core of
radius $\sim (ev)^{-1}$ surrounded by a ``cloud'' of non-Abelian
fields of radius $\sim a$.  Inside this cloud one finds the magnetic
field appropriate to a monopole with both Abelian and non-Abelian
magnetic charges; i.e., the charge appropriate to an isolated massive
$\bbeta$-monopole.  Outside the cloud, only the Abelian charge is
evident.  Thus, the massless $\bgamma$-monopole can be viewed as a
shell of radius $\sim a$ surrounding the massive monopole.  Curiously,
the energy of the solution is independent of $a$, despite the fact
that this parameter is not associated with any symmetry of the system.

We can proceed further and examine the moduli space metric.  Using
the explicit form of the solutions and Eq.~(\ref{gdef}), one
obtains \cite{kwynonabelian}
\begin{equation}
    ds^2 = M d{\bf X}^2 +  {16\pi^2 \over M }d\chi^2 + 
       k \left[{ da^2 \over a} 
      + a \left( d\alpha^2 + \sin^2\alpha d\beta^2  
        + (d \gamma + \cos\alpha d\beta)^2 \right) \right]  
\label{NUSmetric}
\end{equation}
where $k$ is the same constant as in Eq.~(\ref{MSBmetric}).

Now let us return for a minute to the maximally broken case and
imagine approaching the SU(2)$\times$U(1) case by ``rotating'' the
Higgs vector until it is vertical.  This corresponds to taking the
$\bgamma$-monopole mass, and thus the reduced mass $\mu$, to zero.
Taking this limit in the metric of Eq.~(\ref{MSBmetric}) gives
precisely Eq.~(\ref{NUSmetric}), except for a curious change in
notation: The monopole separation $r$ becomes the cloud parameter $a$,
while the spatial angles $\theta$ and $\phi$ combine with the relative
U(1) phase $\psi$ to give the SU(2) Euler angles.

\section{A more complex system}

\def\bhyl{\hat{\bf y}_L}
\def\bhyr{\hat{\bf y}_R}

An example with somewhat more structure is obtained by considering the
case of ${\rm SU}(N+2)$ broken to ${\rm U}(1)\times{\rm
SU}(N)\times{\rm U}(1)$, with the unbroken ${\rm SU}(N)$ lying in
the middle $N\times N$ block in a basis where the eigenvalues of
the Higgs field decrease monotonically along the diagonal.  A purely
Abelian asymptotic magnetic field can be obtained by setting all of
the $\tilde n_j$ and $q_i$ in Eq.~(\ref{NUSgdef}) equal to unity;
i.e., by combining one each of the two species of massive and $N-1$
species of massless monopoles.  The moduli space of such solutions is
$4(N+1)$-dimensional, with the collective coordinates
including a position and U(1) phase for each of the massive monopoles,
a number of SU($N$) orientation angles, and a single cloud parameter,
$b$, that can take on any value greater than or equal to $r$,
the separation between the two massive monopoles.

The solutions can be obtained explicitly \cite{ewyi} by using the Nahm
construction \cite{nahm}.  Although their detailed structure is
somewhat complex, 
their behavior well outside the monopole cores is fairly simple.  The
asymptotic Higgs field can be written as 
\begin{equation}
    \Phi_\infty({\bf r}) = U^{-1}({\bf r}) \,{\rm diag}( v_3, v_2, 
     \dots, v_2,  v_1)\,  U({\bf r}) 
\end{equation}
where $U({\bf r})$ is a gauge transformation whose form will not
concern us and the Higgs eigenvalues satisfy $v_3 > v_2 > v_1$.  The
unbroken SU($N$) lies in the middle $N\times N$ block corresponding to
the repeated eigenvalue $v_2$.
The form of the magnetic field depends on how the
distances $y_L$ and $y_R$ from the two massive monopoles compare to
$b$.   For $y_L, y_R \ll b$,
\begin{equation}
    {\bf B}({\bf r}) = U^{-1}({\bf r})
    \,{\rm diag}\left( {\bhyr\over 2y_R^2}, -{\bhyr\over 2y_R^2},
     {\bhyl\over 2y_L^2}, 
     0,  \dots, 0,  -{\bhyl\over 2y_L^2} \right)  U({\bf r}) + \cdots
\end{equation}
where the dots represent terms that fall off more rapidly with
distance, while for $y \equiv (y_L +y_R)/2 \gg b$
\begin{equation}
    {\bf B}({\bf r}) = U^{-1}({\bf r})
    \,{\rm diag}\left( {\hat{\bf y}\over 2y^2}, 0, 0,
     0,  \dots, 0,  -{\hat{\bf y}\over 2y^2} \right)  U({\bf r})
      + \cdots  \, ,
\end{equation}
Thus, at distances smaller than $b$ one sees fields corresponding to
both Abelian and non-Abelian magnetic charges, but at larger distances
only the Abelian component survives, just as in the SO(5) example.

It is instructive to study the scattering of the two massive monopoles
in these solutions.  Because these monopoles are associated with
orthogonal roots [or, equivalently, because they correspond to
embeddings into commuting $2 \times 2$ blocks of the SU($N$)
matrices], there is no direct interaction between them.  Instead, they
must interact through the massless monopole cloud.  In the low energy
limit, we can use the moduli space approximation to study this process
\cite{xchen}.  The moduli space metric for the maximally broken case
with $N+1$ distinct massive fundamental monopoles is known \cite{kwy}.
The SO(5) example of the previous section suggests that this metric is
still valid when $N-1$ of these monopoles become massless, although,
just as in that case, the physical interpretations of the collective
coordinates change in the massless limit, with the positions of the
massless monopoles being transformed into gauge orientations and a
single gauge invariant parameter, $b$.  The various gauge orientation
and angular variables are most easily handled by expressing them in
terms of the conserved charges and angular momentum.  This still leads
to a fairly complicated dynamics, but matters simplify considerably if
the electric-type charges in the unbroken ${\rm U}(1)\times{\rm
SU}(N)\times{\rm U}(1)$ all vanish.  After separating out the total
center-of-mass motion, one is then left with an effective Lagrangian
\begin{equation}
   L = \left[{\mu\over 2} + {kb \over 4(b^2-r^2)}\right] \dot r^2
        + {kb \over 4(b^2-r^2)}\dot b^2 + {br \over b^2 -r^2}\dot b
        \dot r
        - {b J^2 \over 2r^2(k+\mu b)}
\label{suNlag}
\end{equation}
governing the time development of $b$ and $r$.  Here $J$ is the
angular momentum, $\mu$ is the reduced mass of the massive monopoles
and, as in the previous section, $k$ is a numerical constant.

Analysis of the Euler-Lagrange equations that follow from
Eq.~(\ref{suNlag}) shows that $r$ and $b$ decouple at large times.
Thus, at large $|t|$ the energy of the system is approximately the sum
of a massive monopole kinetic energy $E_r = \mu \dot r^2/2$ and a
cloud energy $E_b =\dot b^2/4b$, with $r$ and $b$ varying
asymptotically as
\begin{eqnarray}
    r &\sim& v|t| + \cdots   \nonumber \\  
    b &\sim& ct^2 + \cdots 
\label{asymbehavior}
\end{eqnarray}
with $v$ and $c$ constants.  In a typical scattering process the cloud
initially decreases in size while simultaneously the two massive
monopoles approach one another.  Eventually $b$ and $r$ reach their
minimum values (generally not at the same time) and then begin to to
increase again according to Eq.~(\ref{asymbehavior}).  In the course
of this process, energy is exchanged between the cloud and the massive
monopoles, so that the final splitting between $E_r$ and $E_b$ at $t
\rightarrow \infty$ is different from the initial one at $t
\rightarrow -\infty$.

One surprising feature of Eq.~(\ref{asymbehavior}) is the quadratic
growth of $b$ with time.  It seems likely that this is an artifact of
the moduli space approximation, and that radiation of massless gluons
will have the effect of reducing this to a linear growth.

\section{Summary and conclusion}

Electric-magnetic duality in gauge theories with the symmetry
spontaneously broken to a non-Abelian subgroup requires the
existence of massless magnetically charged objects that would be the
duals to the electrically charged massless gauge bosons.  These
massless monopoles cannot be exhibited as isolated classical soliton
solutions.  However, as I have shown, their existence is manifested
through clouds of non-Abelian field surrounding one or more massive
monopoles.  These clouds are described by a small number of collective
coordinates, which correspond to the massless monopole degrees of
freedom; for static solutions, the energy is independent of the values
of these parameters.

Duality suggests that these cloud parameters should have counterparts
in the perturbative sector.  It would be instructive to understand more
precisely what these are, and to look in perturbative sector
scattering for analogies with the low-energy scattering of massive
monopoles and massless monopole clouds.  Investigations in this
direction would certainly lead to deeper insights into the properties
of non-Abelian gauge theories.

\vskip .4cm

\noindent I would like to express my appreciation to the organizers of this
conference.  This work was supported in part by the U.S. Department of
Energy.


\vskip .5cm
\begin{thebibliography}{99}
\bibitem{thooft} G 't Hooft, Nucl.  Phys.  {\bf B79}, 276 (1974);
A. M. Polyakov, JEPT Lett. {\bf 20}, 194 (1974).

\bibitem{bps} E.B. Bogomol'nyi, Sov. J. Nucl. Phys. {\bf 24}, 449
(1976); M.K. Prasad and C.M. Sommerfield, Phys. Rev. Lett. {\bf 35},
760 (1975).

\bibitem{montonen} C. Montonen and D. Olive, Phys. Lett. {\bf 72B}, 117
(1977).

\bibitem{osborn}  H. Osborn, Phys. Lett. {\bf 83B}, 321 (1979).

\bibitem{su2index} E.J. Weinberg, Phys. Rev. D {\bf 20}, 936 (1979).

\bibitem{quantization} F. Englert and P. Windey, Phys. Rev. D {\bf
14}, 2728 (1976); P. Goddard, J. Nuyts, and D. Olive, Nucl. Phys. {\bf
B125}, 1 (1997).

\bibitem{Gindex} E.J. Weinberg, Nucl. Phys. {\bf B167}, 500 (1980).

\bibitem{pathologies} A. Abouelsaood, Phys. Lett. {\bf 125B}, 467 (1983);
P. Nelson, Phys. Rev. Lett. {\bf 50}, 939 (1983);
P. Nelson and A. Manohar,  Phys. Rev. Lett. {\bf 50}, 943 
(1983); A. Balachandran, G. Marmo, M. Mukunda, J. Nilsson,
E. Sudarshan, and F. Zaccaria, Phys. Rev. Lett. {\bf 50}, 1553 (1983);
P. Nelson and S. Coleman, Nucl. Phys. {\bf B237}, 1 (1984).


\bibitem{nonabelindex} E.J. Weinberg, Nucl. Phys. {\bf B203}, 445 (1982). 

\bibitem{msa} N.S. Manton, Phys. Lett. {\bf 110B}, 54 (1982).

\bibitem{kwy} K. Lee, E.J. Weinberg, and P. Yi, Phys. Rev. D {\bf
54}, 1633 (1996).

\bibitem{so5} E.J. Weinberg, Phys. Lett. {\bf B119}, 151 (1982).

\bibitem{kwynonabelian} K. Lee, E.J. Weinberg, and P. Yi, Phys. Rev.   
D {\bf 54}, 6351 (1996).

\bibitem{ewyi} E.J. Weinberg and P. Yi, Phys. Rev. D {\bf 58},
046001 (1998).

\bibitem{nahm} W. Nahm, Phys. Lett. {\bf 90B}, 413 (1980); W. Nahm, in
{\it Monopoles in quantum field theory}, N. Craigie et al. eds.
(World Scientific, Singapore, 1982); W. Nahm, in {\it Group
theoretical methods in physics}, G. Denardo et
al. eds. (Springer-Verlag, 1984).

\bibitem{xchen} X. Chen and E.J. Weinberg, in preparation.

\end{thebibliography}
\end{document}